\documentclass[
 reprint,
 amsmath,amssymb,
 aps,
pra,
]{revtex4-2}

\usepackage{graphicx}
\usepackage{physics}
\newtheorem{theorem}{Theorem}
\newtheorem{lemma}{Lemma}

\def\squareforqed{\hbox{\rlap{$\sqcap$}$\sqcup$}}
\def\qed{\ifmmode\squareforqed\else{\unskip\nobreak\hfil
\penalty50\hskip1em\null\nobreak\hfil\squareforqed
\parfillskip=0pt\finalhyphendemerits=0\endgraf}\fi}

\newenvironment{proofof}[1]{\begin{trivlist}\item[]{\flushleft\bf 
Proof of~#1 }}
{\qed\end{trivlist}}

\usepackage{dcolumn}
\usepackage{bm}
\usepackage{hyperref}
\usepackage{enumitem}
\usepackage{colortbl}
\usepackage{tabulary}
\usepackage{etoolbox}
\usepackage{tcolorbox}

\begin{document}

\preprint{APS/123-QED}

\title{Suboptimality of Parity for Distilling  Correlations with Nontrivial Marginals}

\author{Syed Affan Aslam}
 \email{syedaffanaslam@gmail.com}
\author{Areej Ilyas}%
 \email{areej.ilyas@khi.iba.edu.pk}
\author{Jibran Rashid}
 \email{jrashid@iba.edu.pk }
\affiliation{%
 School of Mathematics and Computer Science, \\
Institute of Business Administration, Karachi, Pakistan.
}%




\date{\today}

\begin{abstract}
We prove that the \textsf{PARITY} protocol is optimal for a general class of non-adaptive distillation protocols of all $n$ player nonlocal boxes (\textsf{NLB}s) based on \textsf{XOR} games. The conditional distributions generated by these \textsf{NLB}s are assumed to have trivial local marginals. We also show that already for $n=2$, \textsf{PARITY} is no longer optimal if the local marginals are non-trivial. The \textsf{OR} protocol is shown to perform better and in the process also slightly extend the known correlations that collapse communication complexity. This emphasizes again the need to understand the local properties of nonlocal systems in order to obtain a better characterization of the global behavior. We conclude by showing an equivalence between adaptive distillation protocols that use identical \textsf{NLB}s and \textsf{PARITY} protocol using nonidentical \textsf{NLB}s.
\end{abstract}

\maketitle

\section{Introduction}

Nonlocality serves as a benchmark for genuine quantum behavior. The~research agenda aimed at developing our understanding of nonlocal correlations accessible to physical theories benefits by analyzing correlations produced by nonlocal boxes (\textsf{NLBs})~\cite{Khalfin85, Popescu94b}. These hypothetical models allow investigation of underlying probabilistic frameworks without necessarily identifying complete corresponding physical models. Consequently, they may yield a simpler mathematical model for specific resources and and related analysis for information processing tasks\cite{Barrett05, Buhrman10, Brunner14}. 

We concern ourselves here with concentration of weak multipartite nonlocal correlations through local protocols. Analogous to entanglement distillation, nonlocality distillation protocols allow us to combine multiple copies of a weak correlation source to obtain a stronger one. Developing a better understanding of their properties allows us to rule out theories that allow correlations that do not satisfy \emph{reasonable} information theoretic principles.

H\o yer and Rashid~\cite{Hoyer10} prove optimality of \textsf{PARITY} for distillation for the bipartite case. We generalize the claim by proving that the \textsf{PARITY} protocol is the optimal non-adaptive nonlocality distillation protocol for all $n$ player \textsf{XOR} games. A key limitation in the proof is the assumption that the shared \textsf{NLBs} have trivial local marginals. We show that the \textsf{OR} protocol performs better nonlocality distillation for specific error models with non-trivial marginals. Naik et al.~\cite{Naik23} have identified the \textsf{OR-AND} protocol for a different error model with similar results. We identify parameters for which the \textsf{OR} protocol performs better than any known two copy distillation protocol, non-adaptive or adaptive.

\section{Framework}
\begin{figure}
\centering
\scalebox{.3}{\includegraphics{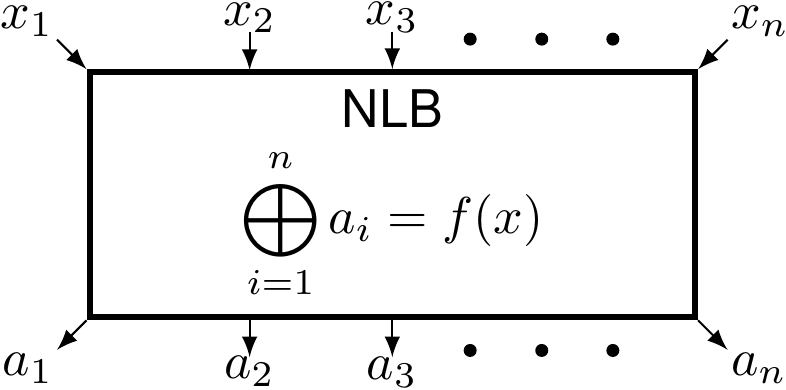}}
\caption{A~multipartite \textsf{NLB} \textsf{N} for \textsf{XOR} games is shared between~$n$ players. On input bit string $x$ the box outputs bit string $a$ such that the parity of $a$ equals $f(x)$ with probability $p(a|x)$.}
\label{fig:mnlb} 
\end{figure}
We consider $n$~player \textsf{XOR} games where players $P_1, P_2, \ldots, P_n$ receive binary inputs $x_1, x_2, \ldots, x_n$ and output bits $a_1, a_2, \ldots, a_n$ respectively. So, the combined length $n$ input $x$ and output $a$ bit strings are elements of $\{0,1\}^n$. The winning condition is satisfied if parity of the output bits equals a boolean function $f$ evaluated on input $x$,~i.e.,
\begin{equation*}\label{cond}
\bigoplus_{i=1}^n a_i = f(x).
\end{equation*}
Each player $P_i$ decides a strategy based on  common knowledge of the fixed function $f$ and their input $x_i$. Given that there is no communication allowed between players after receiving the inputs, the no-signalling conditions limit the probability distributions that players can generate. An \textsf{NLB N} is a resource shared between players that generates conditional probability distributions $p(a|x)$ that satisfy the no-signalling conditions. One way to quantify the strength of the correlation produced by the \textsf{NLB N} is the value $V$ given by
\begin{equation}
\label{mnlbvalue}
V(\textsf{N}) =    \sum_{ f(x)= \bigoplus_i a_i }  p(a|x) - \sum_{ f(x) \neq \bigoplus_i a_i}  p(a|x).
\end{equation}
We parametrize the distributions produced by noisy \textsf{NLBs} on input $x$ by $\delta_x$, such that
\begin{equation}
\label{mnlb}
p(a|x) = 
\begin{cases}
     \frac{1}{2^n}( 1 + \delta_{x}) & \text{if }\oplus_{i=1}^{n} a_i = 0 \textrm{ and}\\
     \frac{1}{2^n} (1 - \delta_{x})  & \text{if }\oplus_{i=1}^{n} a_i = 1.
\end{cases}
\end{equation}
The value $V$ of the \textsf{NLB N} is then given by 
\begin{align}\label{valuerr}
V &= \sum_{a,x \in \{0,1\}^n} (-1)^{\left[ f(x)=\oplus_{i=1}^{n} a_i \right]} p(a|x) \nonumber \\
 &= \sum_{x \in \{0,1\}^n} (-1)^{f(x)} \delta_x,
\end{align}
where the value on input $x$ is $\delta_x$ if $f(x)=0$ and $-\delta_x$ otherwise. Note that the local marginals for each player $P_i$ are trivial,~i.e.,
\begin{equation*}
    p(a_i = 0 | x_i) - p(a_i = 1 | x_i) = 0.
\end{equation*}

\begin{figure}
\centering
\scalebox{.7}{\includegraphics{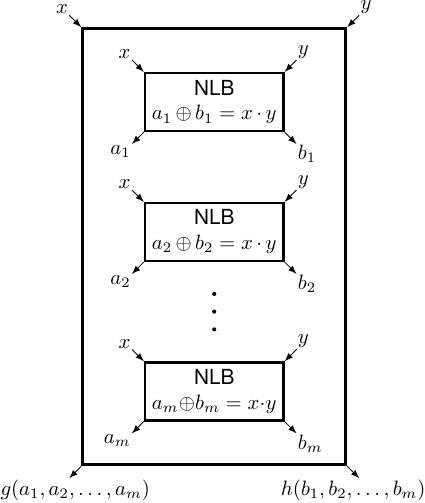}}
\caption{A~non-adaptive distillation protocol for two players using $m$ bipartite \textsf{NLBs}. The \textsf{PARITY} protocol fixes $g$ and $h$ to be the parity function. All boxes receive the same input as the two players.}
\label{fig:nonadaptive} 
\end{figure}

Given $m$ identical \textsf{NLBs} the goal of \textsf{NLB} distillation is to obtain an \textsf{NLB} that achieves as high a value as possible with no communication between the players. H\o yer and Rashid~\cite{Hoyer10} defined a protocol as \emph{non-adaptive} (Figure~\ref{fig:nonadaptive}) if each of the $m$ \textsf{NLBs} takes as input, the original input bits of the players. We consider the same class of protocols here, but note that this is not the most general form of non-adaptive protocols, since the players could locally determine different inputs to the boxes. In our case,  the protocol is completely determined by the boolean functions players use to fix their output bits. The \textsf{PARITY} protocol requires all players to output the parity of the bits received from the $m$ \textsf{NLBs}. Our claim about \textsf{NLBs} of the form in  Equation~\ref{mnlb} is stated in Theorem~\ref{thm:parity}.
\begin{theorem}
\label{thm:parity}
The \textsf{PARITY} protocol is optimal for the class of non-adaptive distillation protocols that provide the same input to all \textsf{NLBs} with trivial marginals. 
\end{theorem}
We also prove Theorem~\ref{thm:or} in Section~\ref{sec:nontriv} that shows that the optimal protocol can change if the local marginals are non-trivial.

\begin{theorem}
\label{thm:or}
The \textsf{OR} protocol can perform better  nonlocality distillation that the \textsf{PARITY} protocol for \textsf{NLBs} with non-trivial marginals. 
\end{theorem}
The proof of optimality for \textsf{PARITY} has free parameters for the expected value for each input while requiring the local marginals to be trivial. The proof for Theorem~\ref{thm:or} is obtained by characterizing the eight dimensional \textsf{NLBs} for the two player CHSH inequality~\cite{CHSH69} in terms of the expected values for each input and  the four local marginals. It is open whether the proof of Theorem~\ref{thm:parity} can be generalized to allow for classification of optimal protocols in terms of local marginals.

\section{Parity is Optimal for \textsf{XOR} Games}
\label{sec:parity}

We prove Theorem~\ref{thm:parity} by first calculating the value achieved by the parity protocol over $m$ \textsf{NLBs} in Lemma~\ref{lm:parityvalue}. 
\begin{lemma}
\label{lm:parityvalue}
The \textsf{PARITY} protocol over~$m$ \textsf{NLBs} of the form in Equation~\ref{mnlb}, attains the value
\begin{equation*}
V=\sum_{x \in \{0,1\}^n} (-1)^{f(x)} \delta_x^m.
\end{equation*}
\end{lemma}

Lemma~\ref{lm:parityvalue} follows from the observation that on input bit string $x$ the \textsf{PARITY} protocol using $m$ \textsf{NLBs}, attains the value $\delta_j^m$ when $f(\mathbf{\hat{x}}) = 0$ and $-\delta_j^m$ otherwise. This is a generalization of Theorem 2 from ~\cite{Forster11}. 

We complete the proof of Theorem~\ref{thm:parity} by proving that the value of all non-adaptive distillation protocols for trivial marginal \textsf{NLBs} is bounded by the value of the \textsf{PARITY} protocol. 
\begin{lemma}
\label{lm:paritybound}
The value of a non-adaptive distillation protocol using at most~$m$ \textsf{NLBs} with the same input, parametrized by Equation~\ref{mnlb}, is upper bounded by the value
\begin{equation*}
\max_{1 \leq k \leq m} \left| \sum_{x \in \{0,1\}^n} (-1)^{f(x)} \delta_x^k \right|.
\end{equation*}
\end{lemma}
\begin{proofof}{Lemma~\ref{lm:paritybound}}
Let $\hat{a}_i = (a_i^1, a_i^2,\ldots, a_i^m)$ be the length $m$ tuple of the output bit string obtained by player $i$ on input bit $x_i$. The combined outputs from the $m$ \textsf{NLBs} received by the $n$ players is given by $\hat{a} = (\hat{a}_1, \hat{a}_2,\ldots, \hat{a}_n)$. Consider the situation where for the input $x = (x_1,x_2,\ldots,x_n)$, all the $n$ players output $\hat{0} = (0,0, \ldots ,0)$ at the end of the protocol. Note that for a fixed input $x$, we can drop the subscript from $\delta$. Let each $\hat{A}_i$ in $\hat{A} =(\hat{A}_1,\hat{A}_2,\ldots,\hat{A}_n)$ represent the set of strings in $\{0,1\}^m$ such that player $i$'s final output at the end of protocol is $0$. Given that the~$n$ players input $x$ into the~$m$  \textsf{NLBs}, the probability that they obtain  output $\hat{a}$ is given by
\begin{align*}    
  p(\hat{a}|x) & =  \prod_{i=1}^{m} \left(\frac{1-\delta}{2^n} + \frac{\delta}{2^{n-1}}\left[\oplus_{j=1}^{n} a_j^i=0\right] \right) \\
& = \left( \frac{1-\delta}{2^n}\right)^m \prod_{i=1}^{m} \left( 1  + \frac{2\delta}{1-\delta}\left[\oplus_{j=1}^{n} a_j^i=0\right] \right) \\
& = \left( \frac{1-\delta}{2^n}\right)^m \left(  \frac{1+\delta}{1- \delta} \right)^{m - |\oplus_{i=1}^n \hat{a}_i|} \\
  & = \frac{1}{2^{m n}} \left( 1 - \delta \right)^{|{\oplus_{i=1}^n \hat{a}_i}|} 
  \left( 1 + \delta \right)^{m - |\oplus_{i=1}^n \hat{a}_i|}.
\end{align*}  

The probability $Q(\hat{A})$ to obtain output $\hat{0}$ can be obtained by summing over all output bit strings $\hat{a}_j \in \hat{A}_j$,~i.e., $Q(\hat{A}(\delta))$ equals
\begin{equation*}
  \sum_{j=1}^n \sum_{\hat{a}_j \in \hat{A}_j} \frac{1}{2^{m n}}\left( 1 - \delta \right)^{\left|\oplus_{j=1}^n \hat{a}_j \right|} \left( 1 + \delta \right)^{m - \left|\oplus_{j=1}^n \hat{a}_j \right|}. 
\end{equation*}
We expand the inner-most product in its $2^m$ terms and express each entry as an evaluation of $\oplus_{j=1}^n \hat{a}_j$ on one of the $2^m$ characters $\chi_z$. Here $\chi_z(\oplus_{j=1}^n \hat{a}_j) = (-1)^{z \cdot (\oplus_{j=1}^n \hat{a}_j) }$ is a character for the group $\mathbb{Z}_2^m$. This gives
\begin{align*}
Q(\hat{A}(\delta)) & = \frac{1}{2^{n m}} \sum_{j=1}^n \sum_{\hat{a}_j\in \hat{A}_j}  \sum_{z \in \{0,1\}^m} \chi_z \left(\oplus_{j=1}^n \hat{a}_j\right) \delta^{|z|}  \\
& =   \sum_{z \in \{0,1\}^m } \delta^{|z|} \prod_{j=1}^{n} \left( \sum_{\hat{a}_j\in \hat{A}_j}   \frac{1}{2^m} \chi_z \left(\hat{a}_j\right) \right) \\ 
& = \sum_{z \in \{0,1\}^m } \delta^{|z|} \prod_{j=1}^{n}  \sum_{s_j} \frac{1}{2^{m}} \chi_z \left(s_j\right)[s_j\in \hat{A}_j].
\end{align*}

We now introduce a set of functions $f_1,f_2, \ldots f_n$ that take the value $+1$ if $s_j \in \hat{A}_j$ and $-1$ otherwise, to express $Q(\hat{A}(\delta))$ as
\begin{align*}
& \hphantom{=} \sum_{z \in \{0,1\}^m } \delta^{|z|} \left( \prod_{j=1}^{n}  \sum_{s_j} \frac{1}{2^{m}} \chi_z \left(s_j\right)\left(
\frac{f_j(s_j) + 1}{2}\right) \right)\\
& = \sum_{z \in \{0,1\}^m } \frac{\delta^{|z|}}{2^n} \prod_{j=1}^{n} \left(\hat{f}_{z}^j + [z = 0] \right),
\end{align*}
where $\hat{f}_{z}^j = \sum_{s_j} \frac{1}{2^m} \chi_z(s_j) f_j(s_j)$ is the Fourier coefficient. Using a similar process we can compute $R(\hat{A}(\delta))$, the probability for the~$n$ players to output even parity. Let $k$ be a bit string in $\{0,1\}^n$ with even parity. We define the $i^{\text{th}}$ element of $\hat{A}^k$ to be the set $\hat{A}_i$ if $k_i = 0$ and its complement $\overline{\hat{A}_i}$ if $k_i=1$, i.e.,~the set containing bit strings on which player $i$'s final output is $1$. For example, if $k$ is equal to the all ones length $n$ bit string $k = \hat{1} = (1,1,\ldots,1)$, then $\hat{A}^k$ corresponds to the set of strings on which the final output of all $n$ players is $1$. Here we equate $Q(\hat{A}^{k=\hat{0}})$ to $Q(\hat{A})$. For a fixed $k$, the corresponding output probability $Q(\hat{A}^k(\delta))$ is given by,
\begin{align*}
  &\hphantom{=} \sum_{z \in \{0,1\}^m } \delta^{|z|} \left( \prod_{j=1}^{n}  \sum_{s_j} \frac{1}{2^{m}} \chi_z \left(s_i\right)\left(
\frac{(-1)^{k_j}f_j(s_j) + 1}{2}\right) \right)\\
&= \sum_{z \in \{0,1\}^m } \frac{\delta^{|z|}}{2^n} \prod_{j=1}^{n} \left( (-1)^{k_j} \hat{f}_{z}^j + [z = 0] \right).
\end{align*}
The probability~$R(\hat{A}(\delta))$ to output even parity bit string is obtained by summing over all $Q$'s, 
\begin{equation*}
R(\hat{A}(\delta)) = \sum_{k} Q(\hat{A}^k(\delta)).
\end{equation*}
Expanding the summation, the middle terms cancel out leaving the expression
\begin{equation*}
R(\hat{A}(\delta)) = \frac{1}{2} \left(1 + \sum_{z \in \{0,1\}^m } \delta^{|z|} \prod_{j}^{n} \hat{f}_{z}^j  \right).
\end{equation*}
The expected value for a fixed $\delta$ can then be expressed as
\begin{equation*}
R(\hat{A}(\delta)) - (1 - R(\hat{A}(\delta)) ) = 2 R(\hat{A}(\delta)) - 1.     
\end{equation*}
Finally, we can determine the bound on the value~$V$ by summing over the biases for each input string $x$ to obtain
\begin{align*}
V & = \sum_{x \in \{0,1\}^n} (-1)^{f(x)} (2R(\delta_x) - 1))\\      
& = \sum_{z \in \{0,1\}^m } \prod_{j=1}^{n}\hat{f}_{z}^j \left(\sum_{x}(-1)^{f(x)} \delta_{x}^{|z|}\right)\\
& \leq \sum_{z \in \{0,1\}^m } \prod_{j=1}^{n}\left| \hat{f}_{z}^j \right| \left(\left| \sum_{x}(-1)^{f(x)}\delta_{x}^{|z|}\right| \right) \\
& \leq \max_{1 \leq k \leq m}\left| \sum_{x}(-1)^{f(x)}\delta_{x}^{k}\right|  \sum_{z \in \{0,1\}^m } \prod_{j=1}^{n}\left| \hat{f}_{z}^j \right| \\
& \leq \max_{1 \leq k \leq m}\left| \sum_{x}(-1)^{f(x)}\delta_{x}^{k}\right|,
\end{align*}
where the last inequality follows from the $\hat{f}_{z}$ being normalized functions.
\end{proofof}
Combining Lemmas~\ref{lm:parityvalue} and \ref{lm:paritybound} proves Theorem~\ref{thm:parity}. Next we restrict out attention to two player \textsf{NLBs} for the CHSH ineuality, but generalize the error model to allow for non-trivial marginals.

\section{Distilling \textsf{NLBs} with Non-Trivial Marginals}
\label{sec:nontriv}

The statement of the \emph{local quantum measurement} principle~\cite{Beigi09} is a proven fact rather than an assumption. It states that in a two player setting if the local systems of Alice and Bob can be described by a quantum system, then their nonlocal correlations also admit a quantum description. The assumption however, lies within the statement. If we assume that the local distributions of Alice and Bob can always be produced by measurements on a quantum system, then according to the conclusion of Barnum et al.~\cite{Beigi09}, their joint distribution also admits a quantum description. The mystery lies in what exactly it entails for the local distributions to be quantum.  

Here we study the influence of local distributions on nonlocality distillation. We generalize the \textsf{NLB} definition to include non-trivial local marginals and then consider them as a resource for nonlocality distillation. For binary input and output, any bipartite no-signalling distribution $p_{ab|xy}$ can be decomposed as
\begin{equation}
\label{eq2}
p_{ab|xy} = \frac{1}{4} \Big( 1 + \left( -1\right)^a E_x + \left( -1\right)^b E_y +\left( -1\right)^{a \oplus b} E_{xy}   \Big),
\end{equation}
where
\begin{equation*}
\begin{aligned}
E_{xy} &= p_{00|xy}+p_{11|xy}-p_{01|xy}-p_{10|xy}, \\
E_{x} &= p_{a=0|x}-p_{a=1|x},  \\
E_{y} &= p_{b=0|y}-p_{b=1|y}. 
\end{aligned}
\end{equation*}
Let the expected value for each input $xy$ be given by 
\begin{equation*}
\delta_1=E_{xy=00},\delta_2= E_{xy=01}, \delta_3 = E_{xy=10} \textrm{ and } \epsilon = E_{xy=11}.
\end{equation*}
Similarly, let the four local marginals be represented by
\begin{equation*}
\alpha=E_{x=0}, \, \beta=E_{x=1}, \, \gamma=E_{y=0} \textrm{ and } \omega=E_{y=1}.  
\end{equation*}
\begin{figure}[t]
\includegraphics{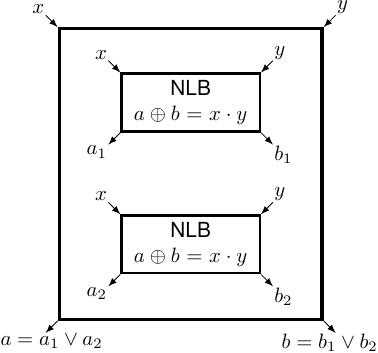}
\caption{\label{nonor} The non-adaptive \textsf{OR} protocol for two \textsf{NLBs}.}
\end{figure}
A local marginal is said to be trivial if it is equal to zero and non-trivial otherwise. The general description of an \textsf{NLB N} with non-trivial marginals is given by
\begin{widetext}
\begin{equation}\label{nlbgen}
N=\frac{1}{4} 
\left( \begin{array}{c@{+}c@{+}c@{+}cc@{+}c@{-}c@{-}cc@{+}c@{-}c@{-}cc@{-}c@{-}c@{+}c}
1&\alpha&\gamma&\delta_1 \,\, & 1&\alpha&\gamma&\delta_1\,\, & 1&\gamma&\alpha&\delta_1 \,\,& 1&\alpha&\gamma&\delta_1 \\ 
1&\alpha&\omega&\delta_2 \,\,& 1&\alpha&\omega&\delta_2 \,\,& 1&\omega&\alpha&\delta_2 \,\,& 1&\alpha&\omega&\delta_2 \\ 
1&\beta&\gamma&\delta_3 \,\,& 1&\beta&\gamma&\delta_3 \,\,& 1&\gamma&\beta&\delta_3 \,\,& 1&\beta&\gamma&\delta_3 \\
1&\beta&\omega&\epsilon \,\,& 1&\beta&\omega&\epsilon \,\,& 1&\omega&\beta&\epsilon \,\, & 1&\beta&\omega&\epsilon \\
\end{array} \right).
\end{equation}
\end{widetext}

Note that the value $V=\delta_1+\delta_2+\delta_3 -\epsilon$ attained by the \textsf{NLB N} in Equation~\ref{nlbgen} is the same as an \textsf{NLB} with trivial marginals. This is a consequence of the fact that the CHSH inequality does not depend on the local marginals. At the same time, we may not choose the values of the local marginals arbitrarily. The value attained by an \textsf{NLB} restricts the values of the marginals in order to ensure no-signalling constraints are not violated. For example, $V=4$ limits the only valid local marginals for a no-signalling \textsf{NLB} to zero. 

Given that non-trivial marginals seem to require more resources to simulate, we consider the possibility that \textsf{NLBs} with non-trivial marginals may have a higher distillation value than those with trivial marginals. Even though the local marginals do not contribute to the value $V$ of a single \textsf{NLB N}, we show in Table~\ref{table:notriv} that via nonlocality distillation protocols an \textsf{NLB} with non-trivial marginals can get distilled to a strictly higher value than the identical \textsf{NLB} with trivial marginals.

This example illustrates that the nonlocal properties of the correlations of an \textsf{NLB} also depend on properties local to Alice and Bob. Distillation provides an operational framework to distinguish \textsf{NLBs} based on the strength of their correlations. Figure~\ref{nonor} gives the \textsf{OR} protocol, which we now examine in terms of its distillation power for non-trivial marginals.

To simpify the analysis we begin by restricting $\delta_1 = \delta_2 = \delta_3 = 1$ and equal but not necessarily trivial local marginals, i.e.,~$\alpha = \beta = \gamma = \omega$. This results in the \emph{correlated} \textsf{NLB} given by
\begin{equation*}
N = \frac{1}{4} 
\left( \begin{array}{c c c c}
2+2\alpha & 0 & 0 & 2-2\alpha \\ 
2+2\alpha & 0 & 0 & 2-2\alpha \\ 
2+2\alpha & 0 & 0 & 2-2\alpha \\
1+2\alpha+\epsilon & 1-\epsilon & 1-\epsilon & 1-2\alpha+\epsilon \\
\end{array}
\right).
\end{equation*}

\begin{figure}[ht]
\includegraphics{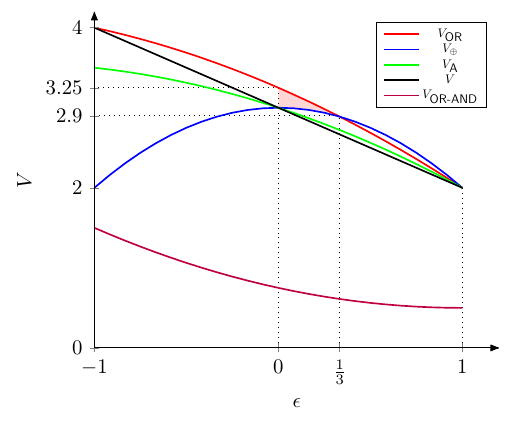}
\caption{\label{plotor} The values for various two copy distillation protocols for $\alpha=\frac{1}{2}$. $V_{\textsf{OR}}$ is the highest distilled value for $\epsilon \in \left[ 0,\frac{1}{3}\right]$. Note that even though the curve for $V_{\textsf{OR}}$ remains at the top for $\epsilon < 0$, these values do not correspond to a valid \textsf{NLB}.}
\end{figure}

\begin{proofof}{Theorem~\ref{thm:or}}
Value $V_{\textsf{OR}}$ of the \textsf{OR} protocol for two copies of \emph{correlated} \textsf{NLBs} is given by,
\begin{equation*} 
\begin{split}
V_{\textsf{OR}} & = (\alpha + \frac{1}{2})(3 - \epsilon ) - \frac{\epsilon ^2}{4} +\frac{5}{4} - 2\alpha \\
 & = \left(\left(\alpha + \frac{1}{2}\right)V\right) + \left(\frac{1}{4} V_{\oplus}\right) + \frac{1}{2} - 2\alpha.
\end{split}
\end{equation*}
It suffices to show that $V_{\textsf{OR}} > \max(V_{\oplus}, V)$, where $V = 3 - \epsilon$ is the value of a single \emph{correlated} \textsf{NLB} and $V_{\oplus} = 3-\epsilon ^2$ is the value of the \textsf{PARITY} protocol on two copies of the \textsf{NLB}. Solving the inequality we obtain the requirement that for 
\begin{equation*}
    \max(1-4\alpha, 2\alpha-1) \leq \epsilon < \frac{4\alpha-1}{3} \textrm{ and } \alpha \in \left[ \frac{1}{4},1 \right],
\end{equation*}
$\textsf{OR}$ protocol peforms better nonlocality distillation than the \textsf{PARITY} protocol.
\end{proofof}
\begin{table}[ht]
\begin{ruledtabular}
 \begin{tabular}{@{}cccr @{.} lr @{.} lr @{.} lr @{.} lr @{.} lr @{.} lr @{.} l@{}} 
$\delta_1$ &  $\delta_2$ &  $\delta_3$& \multicolumn{2}{c}{$\epsilon$}& \multicolumn{2}{c}{$a$} &\multicolumn{2}{c}{$b$} &
\multicolumn{2}{c}{$c$}& \multicolumn{2}{c}{$d$}& \multicolumn{2}{c}{$V$} & \multicolumn{2}{c}{$V_\textsf{A}$}  \\   \hline
$1.00$ & $1.00$ & $1.00$ &$-0$&$70$ & $0$&$01$ &$0$&$01$& $0$&$01$ & $0$&$01$ & $3$&$70$ & $\bf{3}$&$\bf{8360}$  \\
$1.00$ & $1.00$ & $1.00$ &$-0$&$70$ & $0$&$0$ &$0$&$0$& $0$&$0$ & $0$&$0$ & $3$&$70$ & $\bf{3}$&$\bf{8275}$  \\
$0.92$ & $0.92$ & $0.92$ &$-0$&$22$ & $0$&$01$ &$0$&$01$& $0$&$01$ & $0$&$01$ & $2$&$98$ & $\bf{2}$&$\bf{9924}$  \\
$0.92$ & $0.92$ & $0.92$ &$-0$&$22$ & $0$&$0$ &$0$&$0$& $0$&$0$ & $0$&$0$ & $2$&$98$ & $\bf{2}$&$\bf{9867}$  \\
$0.917$ & $0.917$ & $0.917$ &$-0$&$22$ & $0$&$01$ &$0$&$01$& $0$&$01$ & $0$&$01$ & $2$&$971$ & $\bf{2}$&$\bf{97539}$  \\
$0.917$ & $0.917$ & $0.917$ &$-0$&$22$ & $0$&$0$ &$0$&$0$& $0$&$0$ & $0$&$0$ & $2$&$971$ & $\bf{2}$&$\bf{96971}$  \\
\end{tabular}
\end{ruledtabular}
\caption{Column $V$ gives the value attained by the \textsf{NLB} and $V_\textsf{A}$ is the protocol of Allcock et al.~\cite{Allcock09b} for $2$ copies. Note that the distillation value is always higher for the \textsf{NLB} with non-trivial marginals.}
\label{table:notriv}
\end{table}
\begin{figure}[t]
\centering
\includegraphics[width=\linewidth]{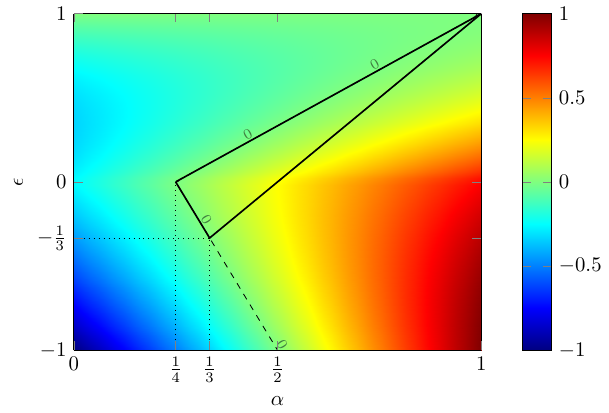}
\caption{The bounded triangle corresponds to the region where \textsf{OR} is optimal among the known protocols for distilling two copies of \emph{correlated} \textsf{NLBs}.}
\label{cont_or}
\end{figure}

Figure~\ref{plotor} shows that the \textsf{OR} protocol performs optimal nonlocality distillation for $\alpha=\frac{1}{2}$ and $\epsilon \in \left[ 0,\frac{1}{3}\right]$. It also outperforms that \emph{adaptive} Allcock et al.~\cite{Allcock09b} protocol. Figure~\ref{cont_or} depicts the entire region in terms of $\epsilon$ and $\alpha$ where the \textsf{OR} protocol is the optimal known distillation protocol for two copies of the \textsf{NLB}. Table \ref{tab:corr} summarizes the behavior of the \textsf{OR} protocol in the optimal distillation region.
 \begin{table}[ht]
 \begin{ruledtabular}
 \begin{tabular}{cc c c c c c c c}
\multicolumn{1}{c}{Range for $\epsilon$} & $\alpha$ & $\delta$ & $\epsilon$  & $V$ & $V_{\oplus}$ & $V_{\textsf{OR}}$& $V_{\textsf{A}}$  \\   \hline
$[-0.04,0.013]$ & $0.26$ & $1.00$ & $0.01$ &  $2.99$& $2.9999$ & $3.00$ & $\bf{3.1112}$ \\
$[-0.20,0.066]$ & $0.30$ & $1.00$ & $0.01$ &  $2.99$& $2.9999$ & $3.04$ & $\bf{3.0914}$ \\
$[-0.30,0.133]$ & $0.35$ & $1.00$ & $0.01$ &  $2.99$& $2.9999$ & $\bf{3.09}$ & $3.0667$ \\
$[-0.20,0.200]$ & $0.40$ & $1.00$ & $0.01$ &  $2.99$& $2.9999$ & $\bf{3.14}$ & $3.0419$ \\
$[-0.10,0.266]$ & $0.45$ & $1.00$ & $0.01$ &  $2.99$& $2.9999$ & $\bf{3.19}$ & $3.0172$ \\
$[\hphantom{-}0.00,0.333]$ & $0.50$ & $1.00$ & $0.01$ &  $2.99$& $2.9999$ & $\bf{3.24}$ & $2.9924$ \\
\end{tabular}
\end{ruledtabular}
\caption{Comparison of the distilled values for \emph{correlated} \textsf{NLBs} attained by various protocols.}
\label{tab:corr}
\end{table}
If we increase the parameter space and allow for $\delta_1 =\delta_2 =\delta_3$, and two local marginals given by $\alpha = \gamma$ and $\beta =\omega$, the values attained for the resulting \emph{symmetric} \textsf{NLBs} are listed in Appendix~\ref{a:sym}. Table~\ref{tab:two} provides a snapshot of how the two non-adaptive protocols behave within this region.

Finally, the known region for trivial communication complexity~\cite{Brassard05} is marginally increased by considering distillation of \textsf{NLBs} with non-trivial marginals. Table~\ref{tab:three} identifies parameter values for \emph{symmetric} \textsf{NLBs} with a single non-trivial marginal. The \textsf{OR} protocol distills these \textsf{NLBs} beyond the required threshold of $4\sqrt{2/3}$.

 \begin{table}[ht]
\begin{ruledtabular}
 \begin{tabular}{c c c c c c c}
$\alpha$ & $\beta$ & $\delta$ & $\epsilon$  & $V$ & $V_{\oplus}$ & $V_{\textsf{OR}}$  \\   \hline
$0.43$ & $0.42$ & $0.99$ & $0.28$ &  $2.69$& $\bf{2.8619}$ & $2.850$  \\
$0.43$ & $0.43$ & $0.99$ & $0.28$ &  $2.69$& $\bf{2.8619}$ & $2.857$  \\
$0.43$ & $0.44$ & $0.99$ & $0.28$ &  $2.69$& $2.8619$ & $\bf{2.864}$  \\
$0.44$ & $0.43$ & $0.99$ & $0.28$ &  $2.69$& $\bf{2.8619}$ & $2.857$  \\
$0.44$ & $0.44$ & $0.99$ & $0.28$ &  $2.69$& $2.8619$ & $\bf{2.864}$  \\
$0.44$ & $0.45$ & $0.99$ & $0.28$ &  $2.69$& $2.8619$ & $\bf{2.871}$  \\
$0.45$ & $0.44$ & $0.99$ & $0.28$ &  $2.69$& $2.8619$ & $\bf{2.864}$  \\
$0.45$ & $0.45$ & $0.99$ & $0.28$ &  $2.69$& $2.8619$ & $\bf{2.871}$  \\
$0.45$ & $0.46$ & $0.99$ & $0.28$ &  $2.69$& $2.8619$ & $\bf{2.878}$  \\
\end{tabular}
\end{ruledtabular}
\caption{Comparison of distilled values via non-adaptive protocols for \emph{symmetric} \textsf{NLBs}.}
\label{tab:two}
\end{table}
 \begin{table}[ht]
 \begin{ruledtabular}
 \begin{tabular}{c c c c c c c}
$\alpha$ & $\delta$ & $\epsilon$  & $V$ & $V_{\oplus}$ & $V_{A}$ & $V_{OR}$  \\   \hline
$0.42$ & $0.99$ & $-0.16$ &  $3.13$ & $2.9744$ & $3.101575$ & $\bf{3.2683}$  \\
$0.41$ & $0.99$ & $-0.18$ &  $3.15$ & $2.9676$ & $3.121425$ & $\bf{3.2735}$  \\
$0.40$ & $0.99$ & $-0.20$ &  $3.17$ & $2.9600$ & $3.141275$ & $\bf{3.2781}$  \\
$0.39$ & $0.99$ & $-0.22$ &  $3.19$ & $2.9516$ & $3.161125$ & $\bf{3.2821}$  \\
$0.38$ & $0.99$ & $-0.24$ &  $3.21$ & $2.9424$ & $3.180975$ & $\bf{3.2855}$  \\
$0.37$ & $0.99$ & $-0.26$ &  $3.23$ & $2.9324$ & $3.200825$ & $\bf{3.2883}$  \\
$0.36$ & $0.99$ & $-0.28$ &  $3.25$ & $2.9216$ & $3.220675$ & $\bf{3.2905}$  \\
\end{tabular}
\end{ruledtabular}
\caption{Choice of parameters that are now included in the region of trivial communication complexity via nonlocality distillation.}
\label{tab:three}
\end{table}

\section{Protocols with Nonidentical Boxes}

\begin{figure}[t]
\centering
\includegraphics[scale=0.14]{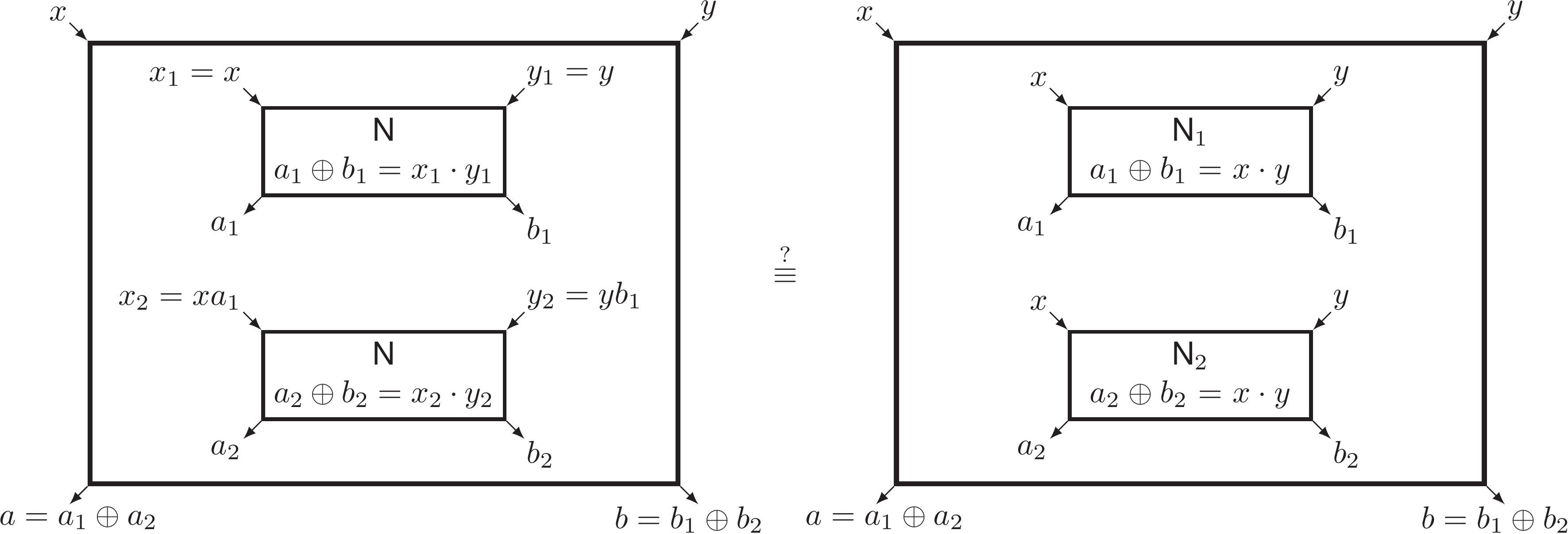}   
\caption{Does there exist an equivalence between the adaptive protocol on the left using two identical \textsf{NLB}s and the non-adaptive protocol on the right using two nonidentical \textsf{NLB}s \textsf{N}$_1$ and \textsf{N}$_2$.}
\label{fig:equiv:brun}
\end{figure}

\begin{figure}[t]
\centering
\includegraphics[scale=0.11]{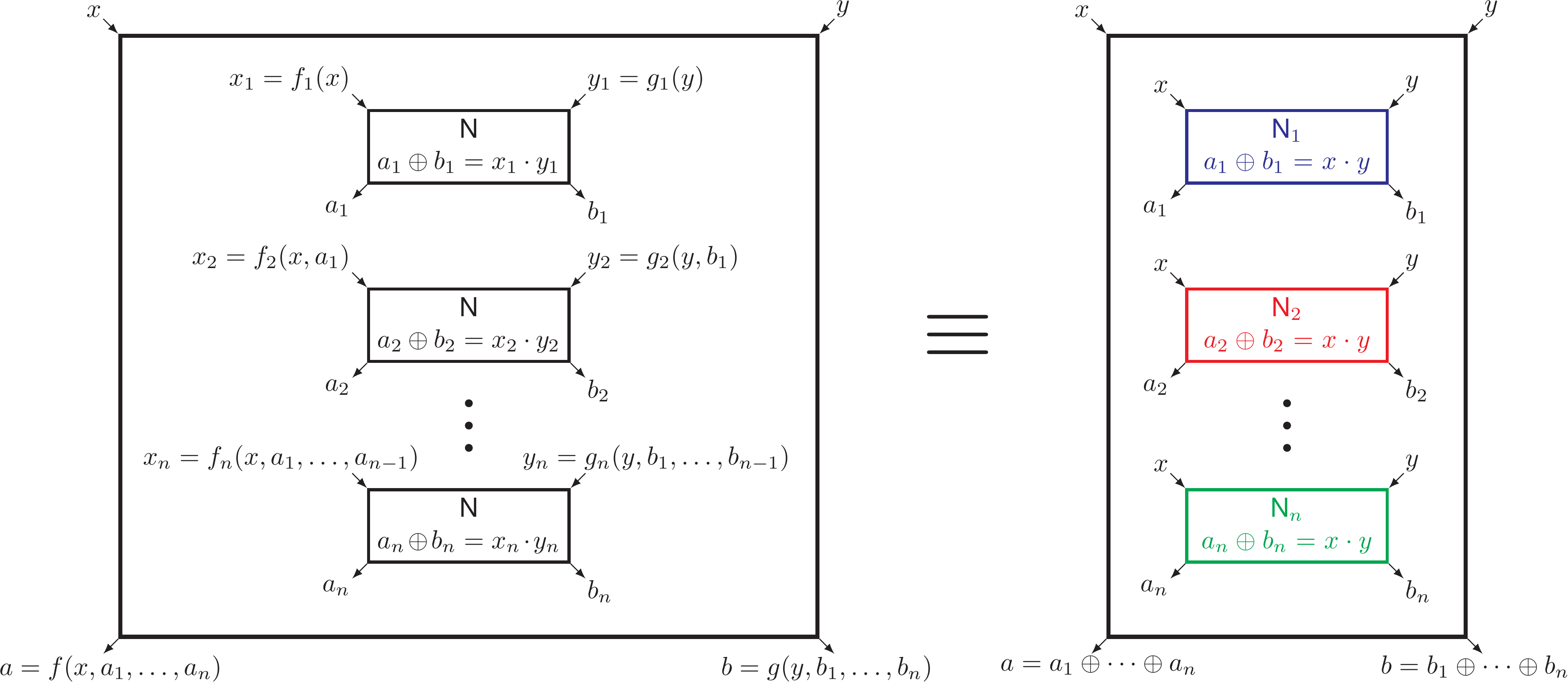}   
\caption{General equivalence between adaptive protocols using identical \textsf{NLB}s and non-adaptive \textsf{PARITY} protocol using nonidentical \textsf{NLB}s.}
\label{fig:equiv:gen}
\end{figure}

Consider whether given an adaptive protocol on~$n$ identical \textsf{NLB}s, we can construct a non-adaptive protocol on~$n$ nonidentical \textsf{NLB}s that simulates the adaptive protocol. Note that we are not asking for a trivial construction where we simply model the final output of the adaptive protocol as another box. We~want to take the adaptiveness of a protocol using identical boxes and push it into the individual boxes. At the expense of having nonidentical boxes we then obtain a non-adaptive protocol that produces the same output as the adaptive protocol. For example, the adaptive protocol of Brunner and Skrzypczyk~\cite{Brunner09} (Figure~\ref{fig:equiv:brun}) maps two isotropic \textsf{NLB}s
\begin{equation*}\label{corrnlb}
\textsf{N} = \frac{1}{4} 
\left( \begin{array}{c c c c}
1+\delta & 1-\delta & 1-\delta & 1+\delta \\ 
1+\delta & 1-\delta & 1-\delta & 1+\delta \\ 
1+\delta & 1-\delta & 1-\delta & 1+\delta \\ 
1-\delta & 1+\delta & 1+\delta & 1-\delta 
\end{array}
\right), 
\end{equation*}
to the \textsf{NLB} given by
\begin{equation*}\label{brun}
\frac{1}{4} 
\left( \begin{array}{c c c c}
1+\delta^2 & 1-\delta^2 & 1-\delta^2 & 1+\delta^2 \\ 
1+\delta^2 & 1-\delta^2 & 1-\delta^2 & 1+\delta^2 \\ 
1+\delta^2 & 1-\delta^2 & 1-\delta^2 & 1+\delta^2 \\ 
\frac{2-\delta^2-\delta}{2} & \frac{2+\delta^2+\delta}{2} & \frac{2+\delta^2+\delta}{2} &\frac{2-\delta^2-\delta}{2}
\end{array}
\right).
\end{equation*}
To construct the equivalent non-adaptive protocol,  for input $xy \in \{00,01,10\}$ we can set the corresponding probability distributions for \textsf{N}$_1$ and \textsf{N}$_2$ equal to $(1+\delta, 1-\delta, 1-\delta, 1+\delta )$. For input $xy=11$, assume that \textsf{N}$_1$ and \textsf{N}$_2$ have expected values $\delta_1$ and $\delta_2$, respectively. Then, the probability that the non-adaptive protocol
outputs $00$ on input $11$ is given by
\begin{equation*}
p_{ab=00|xy=11} = \frac{1-\delta_1 \delta_2}{4}.
\end{equation*}
Equating with the output probability of the adaptive protocol, we obtain
\begin{align*}
 \frac{1-\delta_1 \delta_2}{4} &= \frac{2-\delta^2-\delta}{8} \\
\delta_1 \delta_2 & = \frac{\delta (1+\delta)}{2}.
\end{align*}
Setting $\delta_1 = \delta$ and $\delta_2 = \frac{1+\delta}{2}$ gives the same probability distribution as the adaptive protocol and completes the equivalence construction. In summary, Alice and Bob can non-adaptively use two \textsf{NLBs}, \textsf{N}$_1 =$ \textsf{N} and \textsf{N}$_2$ given by
\begin{equation*}
\textsf{N}_2 = \frac{1}{4} 
\left( \begin{array}{c c c c}
1+\delta & 1-\delta & 1-\delta & 1+\delta \\ 
1+\delta & 1-\delta & 1-\delta & 1+\delta \\ 
1+\delta & 1-\delta & 1-\delta & 1+\delta \\ 
1-\frac{1+\delta}{2} & 1+\frac{1+\delta}{2} & 1+\frac{1+\delta}{2} & 1-\frac{1+\delta}{2} 
\end{array}
\right), 
\end{equation*}
to obtain the same output distribution as the adaptive protocol. Note that both the boxes \textsf{N}$_1$ and \textsf{N}$_2$ have a values less than or equal to the original \textsf{NLB} for all $\delta \geq 0$. We can generalize this construction to establish Theorem~\ref{thm:equiv} (Figure~\ref{fig:equiv:gen}).

\begin{theorem} The output distribution of an adaptive protocol using $n$ identical isotropic \textsf{NLB}s can be simulated by a non-adaptive \textsf{PARITY} protocol over $n$ nonidentical \textsf{NLB}s \textsf{N}$_i$. 
\label{thm:equiv}
\end{theorem}
\begin{proofof}{Theorem~\ref{thm:equiv}}
The probability $p_{00|xy}$ that the adaptive protocol outputs $ab=00$ on input $xy$ is given by a polynomial $q_{xy}(\delta)$ of degree at most $n$, having form
\begin{equation*}
q_{xy}(\delta) = \frac{1}{4^n}\sum_{\substack{f(x,a_1,\ldots,a_n)  = a \\g(y,b_1,\ldots,b_n) =b}} \prod_{j=1}^n (1+(-1)^{[a_j \oplus b_j \neq x_j \cdot y_j]}\delta). 
\end{equation*}
The same probability for a non-adaptive protocol using $n$ nonidentical  \textsf{NLB}s \textsf{N}$_i$ is given by 
\begin{equation*}
\frac{1}{4} \left(1 + \prod_{i=1}^n \delta_{i,xy} (\delta) \right),
\end{equation*}
where $\delta_{i,xy} (\delta)$ is the expected value of the $i^\textrm{th}$ \textsf{NLB} on input $xy$. Equating the two, we obtain
\begin{equation*}
\prod_{i=1}^n \delta_{i,xy} (\delta) = 4 q_{xy}(\delta) -1.
\end{equation*}
Setting each $\delta_{i,xy} (\delta) = c_{i,1} \delta + c_{i,0}$ with the restriction that $\delta_{i,xy} (\delta) \in [-1,1]$ establishes the equivalence.
\end{proofof}

Unfortunately, the proof is not constructive, i.e.~we do not obtain the expressions for each $\delta_{i,xy} (\delta)$. For example, if we can bound the value of each \textsf{N}$_i$ by the value of \textsf{N} we can prove non-distillability of isotropic \textsf{NLB}s in general.

\section{Discussion}

It remains to obtain a complete characterization of optimal non-adaptive distillation protocols in the presence of non-trivial marginals. One possibility is to use the recent algebraic characterization of distillation protocols by Botteron et al.~\cite{Botteron24a,Botteron24b} for this task. Another generalization to consider is the combination of distillation protocols using non-identical \textsf{NLBs}, similar to the approach of Eftaxias, Weilenmann and Colbeck~\cite{Eftaxias23}.

Farkas et al.~\cite{Farkas21} recently showed that nonlocality is not a sufficient condition for the security of device independent QKD protocols. At the same time, Wooltorton, Brown and Colbeck~\cite{Wooltorton24} and Farkas~\cite{Farkas24} have shown that even arbitrarily small nonlocality can be used for DIQKD. This implies that the structure of the nonlocal correlation rather than just the nonlocal value, plays a crucial role in DIQKD protocols. A better understanding of nonlocality distillation in the presence of non-trivial marginals may be useful in identifying the nonlocality requirements for DIQKD.

\bibliography{references}
\newpage
\appendix

\onecolumngrid
\section{Symmetric \textsf{NLBs} with Two Non-Trivial Marginals}\label{a:sym}
We fix $\delta_1 =\delta_2 =\delta_3$, $\alpha = \gamma$ and $\beta =\omega$ in Equation~\ref{nlbgen} to obtain \emph{symmetric} \textsf{NLBs} given by
\begin{equation}\label{ntsym}
N = \frac{1}{4} 
\left( \begin{array}
{c c c c}
1+2\alpha+\delta & 1-\delta & 1-\delta & 1-2\alpha+\delta \\
1+\alpha+\beta+\delta & 1+\alpha-\beta-\delta & 1+\beta-\alpha-\delta & 1-\alpha-\beta+\delta \\ 
1+\alpha+\beta+\delta & 1+\beta-\alpha-\delta & 1+\alpha-\beta-\delta & 1-\alpha-\beta+\delta \\
1+2\beta+\epsilon & 1-\epsilon & 1-\epsilon & 1-2\beta+\epsilon \\
\end{array} \right).
\end{equation}
The values attained by distillation protocols are 
\begin{equation*} 
\begin{split}
V_{\textsf{OR}} &=  \frac{1}{4}(3\delta^2 - \epsilon^2) + \frac{1}{2}(3\delta - \epsilon) + (\beta + 2\delta - 2)\alpha + (\delta - \epsilon)\beta - \frac{1}{2}(\alpha^2 + \beta^2 - 1),\\
V_{\textsf{A}} &= \frac{1}{4} \left( 11\delta^2+2\delta - 2\epsilon \delta - 2\epsilon - \epsilon^2 + (b-a+2d)(\delta - \epsilon)\right) \textrm{ and}\\
V_{\textsf{OR-AND}} &= 2\alpha^2 +\frac{1}{4}\epsilon^2  - \frac{1}{2}\epsilon -\frac{3}{4}\delta ^2 + \frac{3}{2}\delta - \frac{1}{2}.
\end{split}
\end{equation*}

\end{document}